\documentclass{elsart}
\usepackage{feynmp,epsfig}
\def\vec#1{\mathbf #1}
\begin{document}
\begin{frontmatter}
\title{Saturation from nuclear pion dynamics}
\author{M. Lutz$^{a}$, B. Friman$^{a,b}$ and Ch. Appel$^{a}$}
\address{$^{a}$ GSI, Planckstr. 1, D-64291 Darmstadt, Germany}
\address{$^{b}$ Institut f\"ur Kernphysik, TU Darmstadt, D-64289 Darmstadt, Germany}

\begin{abstract}
{We construct an equation-of-state for nuclear matter based on the
chiral Lagrangian. The relevant scales are discussed and an
effective chiral power expansion scheme, which is constructed to
work around the nuclear saturation density, is presented. A
realistic equation-of-state is obtained by adjusting {\em one} free
parameter, when the leading and subleading terms in the expansion
are included. The saturation mechanism is due to correlations
induced by the one-pion-exchange interaction. Furthermore, we find
a substantial deviation from the Fermi-gas estimate of the quark
condensate in nuclear matter already at the saturation density.}
\end{abstract}
\end{frontmatter}


For more than four decades, one of the basic problems of nuclear
physics has been to find a microscopic understanding of the nuclear
equation-of-state in terms of the free nucleon-nucleon
interaction~\cite{Bethe,Day,WirPand}. The present status is that a
quantitative description of nuclear matter can be achieved in a
non-relativistic approach only by invoking a three-body
force~\cite{Ravenhall}, or alternatively with the lowest-order
term in the relativistic Brueckner approach~\cite{BroMach}. A
drawback of the latter, is the lack of a systematic expansion
scheme. Thus, it has so far not been possible to obtain a reliable
estimate for the corrections to the leading term. Such calculations
are based on the traditional semi-phenomenological approach to the
nucleon-nucleon interaction, represented by e.g. the
Urbana-Argonne~\cite{WirPand2} and Bonn~\cite{HolMach} potentials.

During the last few years, a novel approach to the nucleon-nucleon
interaction, based on chiral perturbation theory ($\chi $PT), has
started to emerge. A key element in $\chi $PT is the power
counting, which relies on a separation of scales. This allows one
to organize a systematic approximation scheme. In the nuclear
many-body problem, new scales enter, which requires a modification
of the expansion scheme developed for the nucleon-nucleon
interaction in free space. In this letter we identify the new
scales and present an effective chiral perturbation theory,
appropriate for the nuclear many-body problem. We employ this
scheme to construct a nuclear equation-of-state and compute the
quark condensate in nuclear matter.

The key element of any microscopic theory for the nuclear
equation-of-state is the elementary nucleon-nucleon interaction. In
the context of chiral perturbation theory this problem was first
addressed by Weinberg who proposed to derive a chiral
nucleon-nucleon potential in time ordered perturbation
theory~\cite{Weinberg1,Bira}. An alternative scheme, with chiral
power counting rules applied directly to the nucleon-nucleon
scattering amplitude, was subsequently
proposed~\cite{Lutz1,Lutz3,KSW}. This approach relies on the
crucial observation that the chiral power counting rules can be
generalized for 2-nucleon reducible diagrams. Non-perturbative
effects like the pseudo bound state pole in the $^1S_0$ channel are
generated by properly renormalized local two-nucleon vertices, that
carry an anomalous chiral power $Q^{-1}$.

The counting rules for the vacuum nucleon-nucleon scattering
amplitude provide a suitable starting point for the construction of
chiral power counting rules for the nuclear matter
problem~\cite{Lutz2}. The density expansion, or equivalently the
multiple scattering expansion, is readily combined with the chiral
expansion, once one identifies the Fermi momentum $k_F\sim Q$ as a
further small scale. This identification leads to a systematic
partial resummation of the density expansion, where one avoids the
expansion in large ratios like $k_F/m_\pi$ but exploits the chiral
mass gap and expands in small ratios like $k_F/ m_\chi$ and
$m_\pi/m_\chi $ with $m_\chi \simeq 4\,\pi
\,f_\pi \simeq 1$ GeV.


In the generalized chiral power expansion
scheme~\cite{Lutz1,Lutz3,KSW}, the nucleon-nucleon scattering
amplitude is expanded in the small ratio $Q/\Lambda_L$, but not in
the large one $Q/\Lambda_S$, where $\Lambda_L \simeq m_\chi$ and
$\Lambda_S$ is a small scale, like $
\sqrt{m_N \,\epsilon_D} $ where $\epsilon_D
\simeq 2$ MeV is the binding energy of the deuteron. This leads to
an expansion of the scattering amplitude of the generic form
\begin{equation}
\label{NN}
{\mathcal M}(Q) = \sum_n {\mathcal M}_n\left[{Q\over
\Lambda_S}\right]\left({Q\over \Lambda_L}\right)^n ,
\end{equation}
where the small scale $Q$ is identified with the nucleon momentum
or the mass of the pion $m_\pi$. An immediate consequence of the
chiral expansion schemes, is that the long-range part of the
one-pion-exchange interaction can be treated perturbatively. The
short-ranged or large-momentum part is, as described below,
effectively included in the local part of the chiral Lagrangian,
which is treated non-perturbatively.

When this scheme is applied to the nuclear many-body problem one
finds that the pion dynamics remains perturbative in the sense
discussed above, but the local two-nucleon interaction requires
extensive resummations. This is an immediate consequence of the
chiral power given to the renormalized two-nucleon coupling
$g_R\sim Q^{-1}$. Thus, in nuclear matter at small density, say
$\rho \simeq 0.01$ fm$^{-3}$, it is not sufficient to sum only the
particle-particle ladder diagrams of the local two-nucleon
interaction, as is done in lowest-order Brueckner calculations. We
find that one must sum also the particle-hole and hole-hole ladder
diagrams including all interference terms with self-consistently
dressed nucleon propagators. According to the generalized counting
rules the pions can then be evaluated perturbatively on top of the
parquet resummation for the local two-nucleon interaction,
described above\footnote{The expectation that the nuclear
equation-of-state can be computed microscopically by summing the
parquet diagrams for the local interaction and then including pions
perturbatively may be too optimistic. A refined formulation, which
incorporates the relevant subthreshold singularities of the vacuum
nucleon-nucleon scattering amplitude (see \cite{Lutz3}), may
require a more involved treatment of pionic effects.}. In terms of
vacuum scales this leads to an expansion of the energy per
particle, $E(k_F)$, of the form
\begin{eqnarray}
E(k_F) &=& \sum_n\, E_n\!\!\left[\frac{k_F}{m_\pi},
\frac{k_F}{\Lambda_S}\right]  \left( \frac{k_F}{\Lambda_L}\right)^n .
\label{exp2}
\end{eqnarray}
The expansion coefficients $E_n$ are complicated, presently
unknown, functions of the Fermi momentum $k_F$. However, by
employing appropriate resummations, these functions can be computed
from the free-space chiral Lagrangian~\cite{Lutz:prep}.

Since  the typical small scale $\Lambda_S \simeq 50$ MeV is much
smaller than the Fermi momentum at the saturation density,
$k_F^{(0)} \simeq 265 $ MeV, one may expand the coefficients $E_n$
around $k_F^{(0)} $ in the following manner
\begin{eqnarray}
E_n\!\!\left[\frac{k_F}{m_\pi},
\frac{k_F}{\Lambda_S}\right] &=& E_n\!\!\left[\frac{k_F}{m_\pi},
\frac{k_F^{(0)}}{\Lambda_S}\right]
+\sum_{k=1}^\infty \bar E_n^{(k)}\!\!\left[\frac{k_F}{m_\pi},
\frac{k_F^{(0)}}{\Lambda_S}\right]
\left( \frac{\Lambda_S }{k_F}-\frac{\Lambda_S}{k_F^{(0)}}\right)^k
\; .
\label{exp3}
\end{eqnarray}
Note that we do not expand in the ratio $m_\pi/k_F $. Clearly this
scheme is constructed to work around nuclear saturation density but
will fail at very small densities. However, also other approaches,
like the Walecka model~\cite{Walecka} and the Brueckner scheme, are
not reliable at very low densities.

Physically the expansion (\ref{exp3}) may be interpreted in the
following way. The parquet resummation of the local nucleon-nucleon
interaction gives rise to a density dependence governed by the
scale $\Lambda_S$, while the relevant scale for the pion dynamics
is $m_\pi$. If $\Lambda_S/k_F^{(0)}$ is so small, that the parquet
resummation is close to its high-density limit at normal nuclear
matter density, the expansion (\ref{exp3}) converges rapidly, and
we can, as a first approximation, retain only the leading term in
this expansion. In this case the density dependence of the
coefficients of the expansion (\ref{exp2}) is, to a very good
approximation, given by the pion dynamics.

A systematic derivation of the expansion (\ref{exp2}) and
(\ref{exp3}), applying suitable resummation techniques, will be
presented elsewhere \cite{Lutz:prep}. In this work we pursue a less
microscopic approach. We assume that the coefficients
$E_n(k_F/m_\pi,k_F^{(0)}/\Lambda_S)$ can be computed in an
effective theory whose unknown parameters are adjusted to the
saturation properties of nuclear matter. We employ an effective
Lagrangian density, where the nucleons interact through s-wave
contact interactions and through the exchange of pions. Here we do
not include local p-wave nor three-body interaction terms since
they do not contribute to leading and subleading order in the
chiral expansion. The interaction part of the Lagrangian is given
by
\begin{eqnarray}
{\mathcal L}_{int}(k_F) &=&\frac{g_A}{2\,f_\pi} \, N^\dagger
\left(\vec \sigma \cdot \vec \nabla \right)\Big(
{\vec \pi } \cdot {\vec \tau} \Big)\,N
\nonumber\\
&+& \frac{1}{8\,f_\pi^2}\,\left(g_0(k_F)+\frac{1}{4}\,g_A^2 \right)
\left(  N^\dagger N\right)_1P_{12}^{\,S=1,T=0}
\left( N^\dagger N \right)_2
\nonumber\\
&+& \frac{1}{8\,f_\pi^2}\,\left(g_1(k_F)+\frac{1}{4}\,g_A^2 \right)
\left( N^\dagger N\right)_1P_{12}^{\,S=0,T=1}
\left( N^\dagger N\right)_2
\label{V12}
\end{eqnarray}
where $N$ is the two component nucleon spinor field and
$P_{12}^{\,ST}$ is the projection operator for a two-nucleon state
with spin $S$ and isospin $T$. Note that in the contact interaction
the spatial coordinates of the two particles are identical. The
terms proportional to $g_A^2/4$ are counter terms, which cancel the
local, high momentum, piece of the one-pion-exchange interaction.
Below we discuss the role of the counter terms in more detail.

\begin{figure}[t]
\epsfysize=10.0cm
\begin{center}
\mbox{\epsfbox{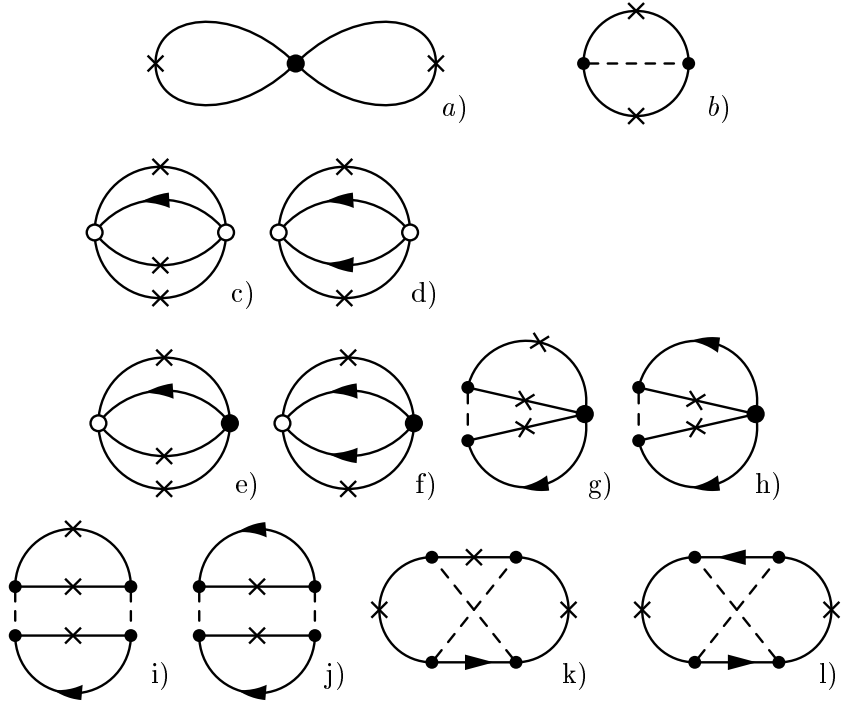}}
\end{center}
\caption{Leading and subleading contributions to the energy of nuclear
matter. The Feyman rules are explained in the text.}
\label{fig1}
\end{figure}

Rather than evaluating the coefficients of the chiral density
expansion (\ref{exp2},\ref{exp3}), we compute the energy directly.
The coefficients $E_n$ can be extracted from the resulting
expressions. In Fig.~\ref{fig1} we show all diagrams of chiral
order $Q^3$ and $Q^4$. They correspond to the leading and
subleading interaction contributions to (\ref{exp2}). The dashed
line is the pion propagator. We split the non-interacting nucleon
propagator into a free-space part
\begin{equation}
\label{prop}
S_N(p_0,\vec p\,) = \frac{1}{p_0-\vec p\,^2/(2 m_N) +
i\varepsilon},
\end{equation}
denoted by a directed solid line, and a density-dependent part
\begin{eqnarray}
\Delta S_N(p_0,\vec p\,) =
2\,\pi\,i\,\delta (p_0-\vec p\,^2/(2\,m_N))\,
\Theta \left(k_F^2-\vec p\,^2\right)\;
\label{cross}
\end{eqnarray}
represented by a line with a 'cross'. Thus, depending on the
diagram, such a line corresponds to a hole line or the Pauli
blocking of a particle line. This separation is useful for the
expansion scheme we adopt.

The pion dynamics which is computed explicitly in our scheme. If
properly renormalized, it is perturbative like in the vacuum case.
The non-perturbative short-range physics is subsumed in the local
interaction. In our scheme, the power counting is much simpler than
in a fully microscopic one, where one would have to sum diagrams
involving the small scales $\Lambda_S$ and $k_F$ to all orders.
This infinite set of diagrams is already included in our
interaction constants $g_0(k_F)$ and $g_1(k_F)$. Consequently, in
order to avoid double counting, diagrams with two or more adjacent
interaction vertices $g_0$ or $g_1$ are forbidden. Therefore we
introduce two types of local 2-nucleon vertices in Fig.~1. The
filled circle represents the full zero-range vertex of (\ref{V12})
proportional to $g_{0,1}+g_A^2/4 $ while the open circle
corresponds to only the counter term proportional to $g_A^2/4 $.
Clearly there are no diagrams of the type e-f in Fig.~\ref{fig1}
with the full contact interaction (filled circles) at both
vertices.

Note, that in our counting $g_{0}$ and $g_{1}$ are of order $Q^0$
since the non-perturbative structures like the deuteron, which give
rise to the anomalously large amplitude $\sim Q^{-1}$ in vacuum,
are dissolved at densities far below the saturation density. We
neglect the density dependence of the effective coupling constants
$g_0$ and $g_1$. This corresponds to retaining only the leading
term in (\ref{exp3}). We can test whether subleading terms are
required, by allowing the coupling constants to be density
dependent.

The contribution of the first diagram in Fig.~\ref{fig1} to the
ground state energy density $\epsilon $ is proportional to
$(g_{0}(k_F)+g_{1}(k_F)) \, k_F^6$. Since the effective vertices
$g_{0}(k_F),g_{1}(k_F) \sim Q^0 $ carry chiral power zero, this
diagram is of chiral order $Q^6 $. The contribution to the energy
per particle $E(k_F)=\epsilon(k_F) /\rho $ is of order $Q^3 $,
since $\rho=2\,k_F^3/(3\,\pi^2)$. In this paper, the chiral power
of a diagram is defined by its contribution to the energy per
particle. The one pion exchange contribution (Fig.~\ref{fig1} b),
is also of chiral order $Q^3$ since it is proportional to $k_F^3 $
times a dimension less function of $k_F/m_\pi$.

The leading interaction contributions to the energy per particle
(Fig.~\ref{fig1} a,b) can be expressed in terms of a momentum
dependent effective scattering amplitude $T_{eff}(p)$:
\begin{eqnarray}
E_T(k_F)&=&-\frac{k_F^3}{2\,\pi^2}\,
\int_0^{1} x^2\,d x\,\Big( 1-x\Big)^2
\Big( 2+x \Big)\,T_{eff}(k_F\,x)
\label{et}
\end{eqnarray}
where
\begin{eqnarray}
T_{eff}(p) &=& \frac{3}{f_\pi^2} \left( g_0+g_1+\frac{g_A^2}{2}\right)
-\frac{3}{2}\,\frac{g_A^2}{f_\pi^2}\,\frac{4\,p^2}{m_\pi^2+4\,p^2}\; .
\label{lead-term}
\end{eqnarray}
In (\ref{lead-term}) the cancellation between the counter terms in
the local effective interaction (\ref{V12}) and the high-momentum
part from the one-pion-exchange interaction is evident. As an
example for the link between the Lagrangian (\ref{V12}) and the
chiral density expansion, we consider the coefficient
$E_3[k_F/m_\pi,k_F^{(0)}/\Lambda_S]$ in (\ref{exp3}). The
contribution of the contact interaction is given by
$E_3/\Lambda_L^3=-2 (g_0 + g_1)/(4
\pi f_\pi)^2$. We note that the small scale $\Lambda_S$ is hidden
in the coupling constants $g_0$ and $g_1$.

Next we consider the diagrams of subleading order, shown in
Fig.~\ref{fig1}. These diagrams fall into two distinct classes:
those with two and three crosses. We do not show the diagrams with
four crosses since their real parts vanish and their imaginary
parts cancel that of the other diagrams so that, to a given order,
the resulting energy is real. Furthermore, we do not include
vacuum-polarization diagrams and vacuum vertex renormalization
diagrams. The former vanish in a non-relativistic
scheme\footnote{The non-relativistic scheme presented here is
identical with the non-relativistic reduction of a relativistic
scheme order by order in the chiral expansion~\cite{Lutz3}.}, while
the latter are suppressed by two chiral powers $Q^2$.

We note that after a proper renormalization, diagrams with only one
cross are implicitly included in the kinetic energy term
\begin{eqnarray}
E_{\rm kin}(k_F) &=&\frac{3}{10}\,\frac{k_F^2}{m_N}\; .
\label{kin-energy}
\end{eqnarray}

The contribution of diagrams d), f), h), j) and l) with two crosses can
again be expressed in terms of an effective scattering amplitude
$T_{eff}(p)$ with
\begin{eqnarray}
T^{(d+f)}_{eff}(p) &=& 6\,\frac{g_A^2}{f_\pi^4}
\left(g_0+g_1+\frac{g_A^2}{4} \right)
\,m_N^2\,I(p)
\nonumber\\
T^{(h)}_{eff}(p)&=&3\left(g_0+g_1+\frac{g_A^2}{2}\right)\frac{g_A^2}{2\,f_\pi^4}\,
\frac{m_N\,m_\pi^2}{8\,\pi\,p}\,i\,
\log \left(1-2\,i\,\frac{p}{m_\pi}\right)
\nonumber\\
&-&6\left(g_0+g_1+\frac{g_A^2}{2}\right)\frac{g_A^2}{f_\pi^4}\,m_N^2\,I(p)
\nonumber\\
T^{(j)}_{eff}(p) &=&\frac{3\,g_A^4}{4\,f_\pi^4}\,
 \frac{m_N\,m_\pi^2}{4\,\pi}\,\left(\frac{1}{m_\pi-2\,i\,p}
-2\,\frac{i}{p}\,
\log \left(1-2\,i\,\frac{p}{m_\pi}\right) \right)
\nonumber\\
&+&6\,\frac{g_A^4}{f_\pi^4}\,m_N^2\,I(p)
\nonumber\\
\nonumber\\
T^{(l)}_{eff}(p) &=&\frac{3\,g_A^4}{2\,f_\pi^4}\,
 \frac{m_N\,m_\pi^2}{4\,\pi}\,\left(
\frac{m_\pi^4+8\,m_\pi^2\,p^2+8\,p^4}
{16\,p\,m_\pi^3(m_\pi^2+2\,p^2)}\,
\Bigg(i\,\log \left(1+4\,\frac{p^2}{m_\pi^2}\right)\right.
\nonumber\\
&+&\arctan \left(
\frac{p}{m_\pi}\,\left(3+4\,\frac{p^2}{m_\pi^2}\right)\right)
-\arctan \left(\frac{p}{m_\pi} \right)
\Bigg)
\nonumber\\
&-&\left.\frac{i\,p}{2\,m_\pi^2}\,
\log \left(1-2\,i\,\frac{p}{m_\pi}\right) \right)
-\frac{3\,g_A^4}{2\,f_\pi^4}\,m_N^2\,\Big(I(p)-2\,I(i\,m_\pi) \Big)\, .
\label{t-eff-Q}
\end{eqnarray}
The loop integral $I(p)$ is divergent and must be regularized:
\begin{eqnarray}
I(p) &=& \frac{1}{8\,\pi^2\,m_N}\,\int_0^\Lambda \,\frac{l^2\,d l}{l^2-p^2-i\,\epsilon}
= \frac{1}{16\,\pi\,m_N} \left(\frac{2}{\pi}\,\Lambda +i\,p
+{\mathcal O}\left( \frac{p^2}{\Lambda^2}\right)\right) \; .
\end{eqnarray}
Here $\Lambda $ is a cutoff parameter.

In (\ref{t-eff-Q}) the appropriate symmetry factors are included.
We note that the diagrams d), f), h), j) and l) in Fig. \ref{fig1} are
all divergent. However, the sum of all diagrams is finite once we
include a further counter term in the spin triplet channel:
\begin{eqnarray}
g_0 \rightarrow g_0
-\frac{m_N\,\Lambda}{4\,\pi^2}\,\frac{g_A^4}{f_\pi^2}
\;.
\end{eqnarray}
This counter term, which is formally of order $Q$, cancels the
divergence due to the second-order tensor interaction. Collecting
all real terms from (\ref{lead-term}) and (\ref{t-eff-Q}) we find
\begin{eqnarray}
T_{eff}(p) &=& \frac{3}{f_\pi^2}\,\Big(g_0+g_1 \Big)
+\frac{3}{2}\,\frac{g_A^2}{f_\pi^2}\,\frac{m_\pi^2}{m_\pi^2+4\,p^2}
\nonumber\\
&+&\frac{3\,g_A^2}{f_\pi^2}\,
\Big(g_0+g_1\Big)\,\frac{m_\pi\,m_N}{16\pi
\,f_\pi^2}\, \frac{m_\pi}{p}\, \arctan \left( \frac{2\,p}{m_\pi}\right)
\nonumber\\
&+&\frac{3\,g_A^4}{f_\pi^2}\,\frac{m_\pi\,m_N}{16\pi
\,f_\pi^2}\,\Bigg( \frac{m_\pi^2}{m_\pi^2+4\,p^2}
-\frac{3}{2}\,\frac{m_\pi}{p}\, \arctan \left( \frac{2\,p}{m_\pi}\right)
\Bigg)
\nonumber\\
&+&\frac{3\,g_A^4}{f_\pi^2}\,\frac{m_\pi\,m_N}{16\pi\,f_\pi^2}
\Bigg(-
\frac{p}{m_\pi}\, \arctan \left( \frac{2\,p}{m_\pi}\right)-1
\nonumber\\
&+&\frac{m_\pi^4+8\,m_\pi^2\,p^2+8\,p^4}
{8\,p\,m_\pi\,(m_\pi^2+2\,p^2)}\,
\Bigg(\arctan \left(
\frac{p}{m_\pi}\,\left(3+4\,\frac{p^2}{m_\pi^2}\right)\right)
\nonumber\\
&-&\arctan \left(\frac{p}{m_\pi} \right)
\Bigg)
\Bigg)\, .
\label{final-t}
\end{eqnarray}
The corresponding contribution to the energy per particle is again
given by Eq.~(\ref{et}).

We now turn to the remaining diagrams c), e), g), i) and k). Their
contribution to the energy per particle reads
\begin{eqnarray}
E_P(k_F) &=&\frac{m_N\,k_F^4}{(4\,\pi\,f_\pi)^4}\left(
\frac{3}{2}\,g_A^2\left(g_0+g_1+\frac{g_A^2}{4} \right)\,J_0
\right.
\nonumber\\
&-&\left.
\frac{3}{2}\,g_A^2 \,\left(g_0+g_1+\frac{g_A^2}{2} \right)J_1(k_F)
+\frac{3}{2}\,g_A^4\,J_2(k_F)-\frac{3}{8}\,g_A^4\,J_3(k_F)
\right)
\end{eqnarray}
where
\begin{eqnarray}
J_{n<3}(k_F) &=&\frac{24}{k_F^7}
\int_0^{k_F} \vec q^2\,d\vec q
\int_{-1}^{1}d\,x_{\vec q}
\left(k_F\,|\vec q|\,x_{\vec q} +\frac{k_F^2-\vec q^2\,x^2_{\vec
q}}{2}
\,\log \frac{k_F+|\vec q|\,x_{\vec q}}{k_F-|\vec q|\,x_{\vec q}}
\right)
\nonumber\\
&&\;\;\;\;\;\;\;\;\;\;\;\;\;\;\;\;\;\;\;\;\;\;\;\;\;\;\;\;\;\;
\cdot \Big(F_n(0)- F_n(\lambda (\vec q,x_{\vec q}))\Big)
\nonumber\\
J_3(k_F) &=&\frac{24}{k_F^7}\int_0^{k_F} \vec p^2\,d\vec p
\int_{-1}^{1}x_{\vec p}\,d\,x_{\vec p}
\int_{-1}^{1}x_{\vec q}\,d\,x_{\vec q}
\left(
\frac{\Theta \left(x_{\vec q}^2+x_{\vec p}^2-1\right)}
{\sqrt{x_{\vec q}^2\,x_{\vec p}^2\left(x_{\vec q}^2+x_{\vec
p}^2-1\right)}}-2\right)
\nonumber\\
&&\;\;\;\;\;\;\;\;\cdot \Big(F_1(\lambda(\vec p,x_{\vec
q}))-F_1(0)
\Big)\,\Big(F_1(\lambda(\vec p,x_{\vec p}))-F_1(0)
\Big)
\end{eqnarray}
and
\begin{eqnarray}
\lambda (\vec q,x_{\vec q} ) &=&\sqrt{k_F^2-\vec q^2 (1-x_{\vec q}^2)}
-|\vec q|\,x_{\vec q}
\nonumber\\
F_0(l)&=&\frac{\vec l^2}{2}
\nonumber\\
F_1(l)&=&\frac{\vec l^2}{2}-\frac{m_\pi^2}{2}
\,\log \left(\vec l^2+m_\pi^2\right)
\nonumber\\
F_2(l) &=&\frac{\vec l^2}{2}
-\frac{m_\pi^4}{2\,(\vec l^2+m_\pi^2)}
-m_\pi^2\,\log  \left(\vec l^2+m_\pi^2\right)\, .
\end{eqnarray}
The integral $J_0$ can be performed analytically with the result
$J_0= 16\,(11-\ln 4)/35$.

\begin{figure}[t]
\epsfysize=8.0cm
\begin{center}
\mbox{\epsfbox{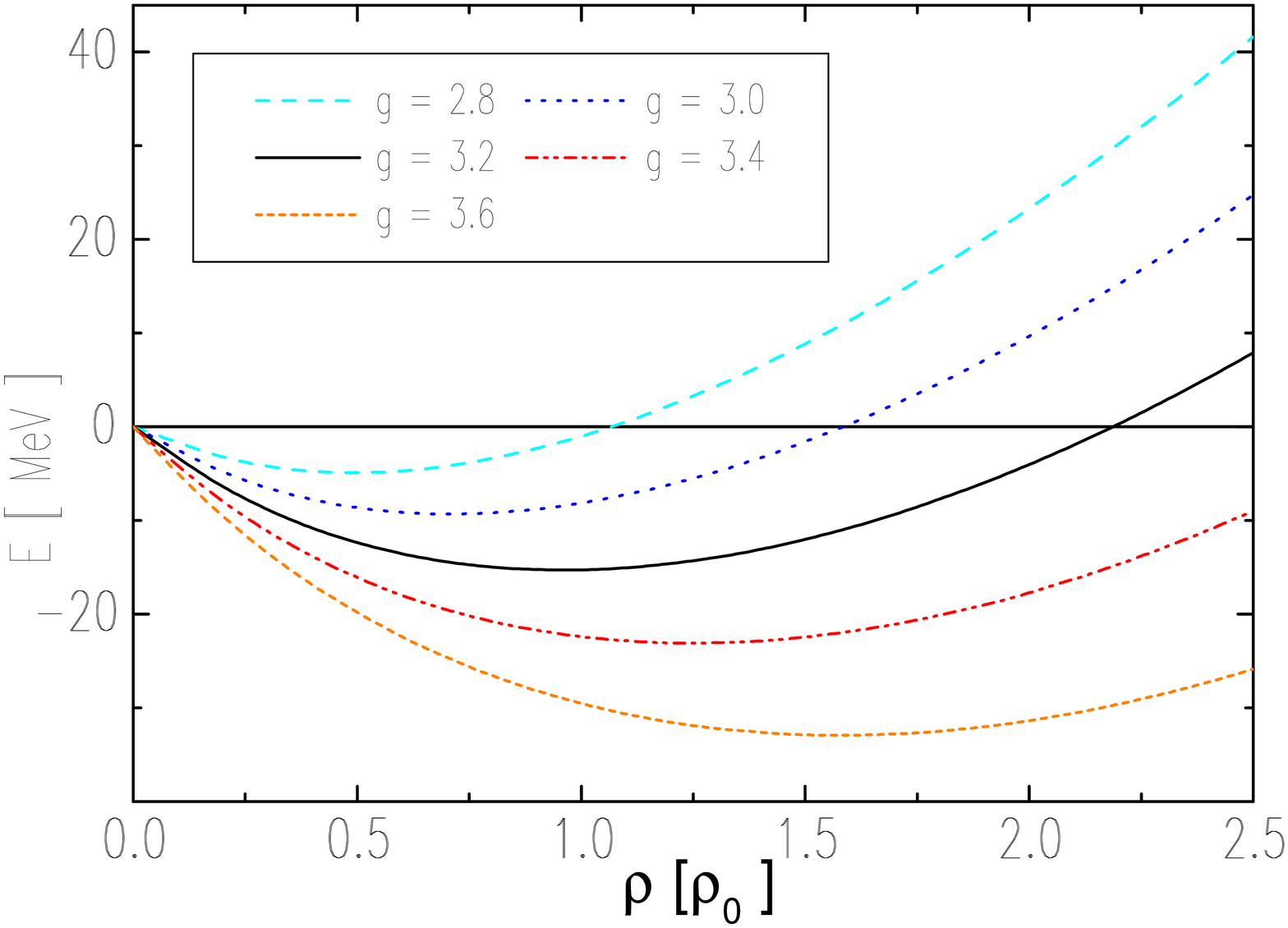}}
\end{center}
\caption{The equation-of-state for isospin-symmetric nuclear matter.}
\label{fig2}
\end{figure}

The resulting equation-of-state is given by $E(k_F) = E_{kin} + E_T
+ E_P$. We note that in symmetric nuclear matter there is
effectively only one free parameter, namely the effective coupling
constant $g=g_0 + g_1$, since we neglect the density dependence of
$g_0$ and $g_1$, as discussed above. In Fig.~\ref{fig2} we show the
energy per nucleon for $g=2.8, 3.0, 3.2, 3.4$ and $3.6$. We
emphasize that the coupling functions $g_{0}(k_F),g_{1}(k_F)$ are
to be determined from the nuclear equation-of-state. Thus, by
allowing $g_{0}(k_F)$ and $g_{1}(k_F)$ to be arbitrary functions of
$k_F$ we could have trivially obtained a realistic
equation-of-state. However there is a strong consistency
constraint: according to our scale argument discussed above, the
density dependence of the coupling functions must be weak for $k_F$
larger than the small scales, which formally have been integrated
out. Thus, our scheme would have to be rejected, had we found a
strong density dependence of $g_0$ and $g_1$. In fact, the density
independent set of parameters $g \simeq 3.23 $, $g_A \simeq 1.26 $,
$m_\pi \simeq 140 $ MeV  and $f_\pi\simeq 93 $ MeV yields an
excellent equation-of-state. We obtain saturation at the Fermi
momentum $k_F^{(0)} \simeq 265 $ MeV, which corresponds to a
density $\rho_0 \simeq 0.16$ fm$^{-3}$, in agreement with the
empirical value. Furthermore, the empirical binding energy of 16
MeV is reproduced and we obtain an incompressibility of $\kappa \simeq 
218 $ MeV, compatible with the empirical value $(210 \pm 30) $ MeV
of ref.~\cite{Blaizot}.

It is interesting to explore the convergence properties of the
expansion (\ref{exp2}). At the saturation density, the leading term
of chiral order $Q^2$ (the kinetic energy) contributes $22.5$ MeV
to the energy per particle $E(k_F)$, the terms of order $Q^3$
(Fig.~1 a-b) contribute $-93.5$ MeV and the terms of order $Q^4$
(Fig.~1 c-l) $+55.0$ MeV. We observe a partial cancellation between
the $Q^3$ and $Q^4$ terms. However, based on these terms alone, one
can draw only qualitative conclusions on the convergence properties
of the expansion. Quantitative results have to await the evaluation
of terms of higher order in $Q$. As one finds e.g. for the
hole-line expansion in Brueckner theory~\cite{Day}, it may well be
that the convergence is in fact much better than suggested by the
first few (fairly large) terms.

The saturation mechanism of our model is quite different from that
of popular effective models, like the Skyrme and Walecka models. In
the Skyrme model the saturation is due to the density and velocity
dependence of the zero-range effective interaction~\cite{Blaizot},
while in the Walecka model it is a relativistic
effect~\cite{Walecka}. In neither of these models are pionic
degrees of freedom explicitly included. In our model, on the other
hand, saturation is obviously due to pion dynamics, since for $g_A
= 0$ only the diagram in Fig.~1a survives and $g_0$ and $g_1$ are
independent of density. This is reminiscent of the saturation
mechanism in the Brueckner approach, which, to a large extent, is
due to the second-order tensor contribution~\cite{Bethe}. We note
however that there is no one-to-one correspondence between the
pion-exchange diagrams in the two approaches. For instance, the
short-range part of the iterated one-pion-exchange is in our
approach, through the renormalization procedure, subsumed in the
zero-range interaction. Since the coupling constants $g_0$ and
$g_1$ are adjusted to reproduce the properties of nuclear matter,
this part of the one-pion-exchange interaction is effectively
treated non-perturbatively. In the Brueckner approach, on the other
hand, the divergencies are regularized by form factors.
Consequently, there the corresponding finite contributions to the
energy are associated with the original pion-exchange diagrams.
Since the one-pion-exchange is perturbative only once the
short-range part has been removed, it must be iterated to all
orders in the Brueckner approach.


Our equation-of-state is useful for applications, where the pion
dynamics plays an important role. As an example, we consider the
quark condensate in nuclear matter. The quark condensate, $\langle
\bar qq \rangle$, is an order parameter for the spontaneously broken chiral
symmetry. Thus, its dependence on baryon density, indicates to what
extent chiral symmetry is restored in nuclear matter. Furthermore,
it is an important ingredient in the QCD sum rules \cite{S.H.Lee}
and in the Brown-Rho scaling approach~\cite{BR-scaling}. According
to the Feynman-Hellman theorem the quark condensate can be
extracted unambiguously from the total energy per particle, $m_N +
E(\rho) $, of nuclear matter, once its dependence on the current
quark mass is known:
\begin{eqnarray}
\langle \bar qq \rangle (\rho ) - \langle \bar qq \rangle (0)
&=& \rho \,\frac{\partial}{\partial\,m_Q} \, \left(m_N + E(\rho,
m_Q )\right)\,
.
\label{FH1}
\end{eqnarray}
Here $m_Q=(m_u+m_d)/2$ is the average of the $u$ and $d$ quark
masses. In our approach the contribution of the pion degrees of
freedom to the energy is included in a chirally consistent manner
to order $Q^4$. This allows us to estimate the effect of
correlations on the quark condensate in nuclear matter to leading
and subleading order.

It is convenient to consider the relative change of the quark
condensate
\begin{eqnarray}
\frac{\langle \bar q\, q \rangle (\rho)}
{\langle \bar q\, q \rangle (0)}
=1-\frac{\Sigma_{\pi N}\,\rho}{m_\pi^2\,f_\pi^2}
-\frac{\alpha_\pi(\rho)\,\rho}{2\,m_\pi\,f_\pi^2}
\label{FH2} \; .
\end{eqnarray}
The penultimate term in (\ref{FH2}) is the modification of the
quark condensate in a free Fermi gas of nucleons, where
\begin{eqnarray}
\Sigma_{\pi N} = m_Q \,\langle N|\bar q\, q |N\rangle
=m_Q\,\frac{d\,m_N}{d\,m_Q} \;
\label{sigma-term}
\end{eqnarray}
is the pion-nucleon sigma term. Finally, the last term in
(\ref{FH2}) is due to the dependence of the interaction energy on
the pion mass
\begin{eqnarray}
\alpha_\pi(\rho) &=&
-\frac{2\,m_\pi\,f_\pi^2}{\langle \bar q\, q \rangle (0)}
\,\frac{\partial }{\partial\, m_Q}\, E(\rho ,m_Q )
=\left(1+{\mathcal
O}\left( m_\pi^2 \right) \right)\frac{\partial }{\partial
\,m_\pi}\, E(\rho, m_\pi )
\label{alpha}
\end{eqnarray}
where we use the Gell-Mann--Oakes--Renner relation to convert the
dependence on the current quark mass $m_Q$ to a dependence on the
pion mass. We emphasize that the second term on the right-hand side
of (\ref{FH2}), which was first written down in
refs.~\cite{Drukarev,Lutz92,Cohen}, does not probe the interactions
in nuclear matter. Therefore the linear density dependence of the
quark condensate, which results when only this term is retained,
should be considered with caution. It is a priori not clear that
this term is the most important one at the saturation density and
beyond.

\begin{figure}[t]
\epsfysize=8cm
\begin{center}
\mbox{\epsfbox{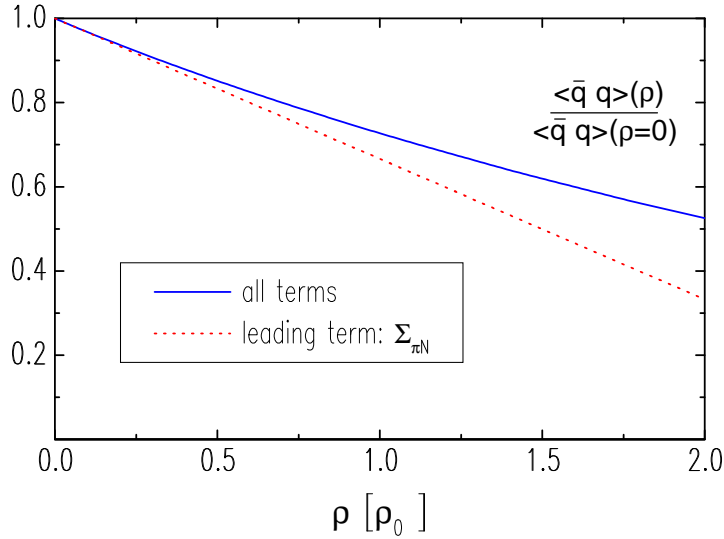}}
\end{center}
\caption{The quark condensate in isospin symmetric nuclear matter.}
\label{fig3}
\end{figure}

Note that here we neglect any implicit $m_\pi$ dependence of the
effective couplings $g_{0}$ and $g_{1}$, since naive counting
arguments suggest that the corresponding contribution to the energy
per particle would be of order $Q^5$. In Fig.~\ref{fig3} we show
the resulting quark condensate in nuclear matter. We confront the
'leading' term, which is proportional to the sigma term
$\Sigma_{\pi N}\simeq 45 $ MeV, with the full result given by
(\ref{FH2}). The inclusion of pionic interaction effects
counteracts the reduction of the condensate due to the leading
term. Our results confirm calculations performed within the
Brueckner \cite{Li} and Dirac-Brueckner~\cite{Brockmann} approach
qualitatively insofar that the nuclear many-body system reacts
against chiral symmetry restoration. This result has important
implications for the restoration of chiral symmetry in matter at
finite baryon density.

\end{document}